\input{aipcheck}

\documentclass[
    ,final            
  ]
  {aipproc}

\layoutstyle{6x9}

\begin{document}

\title{Non-abelian black holes and black strings in higher dimensions}

\classification{\texttt{11.10.Kk, 03.50.Kk, 04.40.Nr, 04.50.Gh}}
\author{Betti Hartmann}{
  address={School of Engineering and Science, Jacobs University Bremen, 28759 Bremen, Germany}
}

\begin{abstract}
We review the properties of static,
higher dimensional black hole solutions in theories where
non-abelian gauge fields are minimally coupled to gravity. It is shown that 
black holes with hyperspherically symmetric horizon topology do not exist in $d > 4$, but
that hyperspherically symmetric black holes can be constructed numerically in generalized
Einstein-Yang-Mills models.
5-dimensional black strings with horizon topology $S^2 x S^1$ are also discussed. These are so-called
undeformed and deformed non-abelian black strings, which are translationally invariant and correspond to
4-dimensional non-abelian black holes trivially extended
into one extra dimensions. The fact that black strings can be deformed, i.e. axially symmetric for constant values of the extra coordinate is a new feature as compared to black string solutions of Einstein (--Maxwell) theory. It is argued that these non-abelian black strings are thermodynamically unstable.

\end{abstract}

\maketitle


\section{Introduction}
Originally motivated by recent results in string theory, brane world models assume that our universe is a $(3+1)$-dimensional
submanifold (a ``3-brane'') embedded in a higher dimensional space-time.
While in (super)string theory - which is only consistent in $10$ dimensions
- the extra dimensions are compactified on circles with radii of Planck length,
brane world models have large or even infinite extra dimensions. Black holes in
higher dimensions are of interest for these scenarios since gravity - unlike the Standard model
fields, which are localized on the brane - ``lives'' in all dimensions. 
Moreover, since the fundamental, higher-dimensional
gravity scale could be as low as the TeV scale (which would solve the hierarchy problem)
the production of small black holes in future collider experiments (e.g. at LHC)
as well as in high energy cosmic rays is a possibility. Intriguingly,
black holes in higher dimensions are not by far as well understood as their counterparts
in 4 dimensions. An important point is that black holes in higher dimensions
are not restricted to (hyper)spherically symmetric
horizon topologies. If the space-time is asymptotically flat, the only {\it static} black holes
are the Tangherlini-Schwarzschild and Tangherlini--Reissner-Nordstr\"om solutions
in the d-dimensional vacuum Einstein, respectively Einstein-Maxwell model \cite{tan}. 
These black hole solutions have horizon topology $S^{d-2}$. 
Uniqueness theorems for static black holes in asymptotically flat higher dimensional space-times 
have been well established. The uniqueness of the Tangherlini-Schwarzschild solutions in the vacuum
Einstein model, respectively of the Tangherlini--Reissner-Nordstr\"om 
solutions in the Einstein-Maxwell model in the
non-degenerate case has been proved \cite{gis}. In the degenerate case,
the higher dimensional analogues of the Majundar-Papapetrou solutions \cite{remark,pm} 
were shown to be the unique solutions  of the Einstein-Maxwell system \cite{rogatko3}.
This was extended to theories with other matter sources.
The uniqueness of the Gibbons-Maeda solutions \cite{gm}
in Einstein-Maxwell-dilaton theories has been proved 
\cite{gis2}. Uniqueness theorems for static black hole solutions
also exist for a gravitating $\sigma$ model \cite{rogatko} 
as well as for electrically and magnetically charged black holes
in generalised Einstein-Maxwell theories (these theories contain
a $(d-2)$-gauge form) \cite{rogatko2}. 

If, however, {\it static} solutions in asymptotically $(d-\tilde{d})$-
dimensional Minkowski space $M^{d-\tilde{d}}$
times a compact Ricci-flat $\tilde{d}$-dimensional manifold  $X^{\tilde{d}}$
(the simplest and most studied case being $M^4\times S^1$) are considered, one would expect 
black objects with different horizon topologies - depending
on the size of the extra dimensions. When the size of the compact manifold is large (resp. small) in comparison to the horizon of the black object,
one would expect the object to be a black hole with horizon topology $S^{d-2}$ 
(black string with horizon topology $S^{d-\tilde{d}-2}\times X^{\tilde{d}}$). The phase diagram of 
{\it static} black objects arising as solutions
in 5-dimensional vacuum $M^4\times S^1$ have been studied in great detail \cite{kol,wiseman,wiseman2,obers,ho}. For small and large horizon sizes
(compared to the size of the extra dimension),
there are black hole solutions with horizon topology $S^3$ (sometimes also called ``caged black holes''), respectively
the black string solutions. Caged black holes have been studied both analytically
\cite{cbh_analytic} as well as numerically  
\cite{kol,kw2,kw}.
The black string is a simple product of a 
$4$-dimensional
Schwarzschild black hole with a circle $S^1$ and has 
thus horizon topology $S^2\times S^1$.
As shown by Gregory and Laflamme \cite{gl}, the $5$-dimensional black string is unstable to linear perturbations
if the length of the extra compact dimension is larger than (roughly) the horizon radius of the
$4$-dimensional Schwarzschild black hole. (Note that this instability
also exists for black $p$-branes, i.e. objects with horizon topology
$(d-p)$-dimensional Schwarzschild $\times S^p$.) The horizon starts oscillating in the extra coordinate direction
and the black string decays to a black hole. In the transition region so-called ``non-uniform'' black strings
exist. These solutions have metric functions depending on the extra-dimensional coordinate
and have been constructed numerically using a relaxation method \cite{wiseman,wiseman2}. 
Recently, there has been some debate whether the original
assumption of Gregory and Laflamme that the end of the instability of
the black strings is a black hole was correct \cite{kol,horowitz}. 
It was argued that the end result should instead be a stable non-uniform black
string \cite{horowitz}. The problem is still not fully solved, however numerical
results give good hints that the scenario originally proposed by Gregory and Laflamme
is correct \cite{wiseman,kw}.
 
As shown in \cite{ho,hm}, the charged black strings
are quite different from their uncharged counterparts (at least in the (near-) extremal limit).
It was shown that an arbitrarily charged black hole can fit into
a space-time with the compactified direction being arbitrarily small. Moreover,
close to the extremal limit uniform and non-uniform {\it stable} black strings (and in more generality black $p$-branes)
were shown to exist with the same mass and charge. The important point here is that - different from the uncharged
case - the non-uniform solutions don't result from the instability of the uniform
solution (The instability of the uniform solutions has been proved in \cite{gl2}). 

In this paper, we want to discuss higher dimensional static black hole solutions
that carry non-abelian gauge fields. The interplay between non-abelian gauge fields and gravity in 4 space-times dimensions leads to the existence of so-called  non-abelian black holes,
which violate the No-hair conjecture since they carry non-trivial matter fields 
outside their regular horizon \cite{coloured}. Moreover, it was shown by explicit construction
that static, axially symmetric non-abelian black holes exist \cite{kk} in 4 dimensions, thus violating Israel's theorem.
If one extends the model to more than 4 space-time dimensions, it can be shown that
hyperspherically symmetric static non-abelian black holes do not exist in $d > 4$.
This was shown previously for $d=5$ \cite{bhr} and is extended to the case $d >5$ in this paper.
There are two possibilities to circumvent this problem, which are both discussed here.
The first possibility is to add additional terms from the gravity or gauge field hierarchy.
Then hyperspherically symmetric solutions are possible \cite{bct, bcht}. The second possibility is to consider solutions with other than spherical horizon topology, e.g. black strings \cite{h1,bh1,bhr}.

\section{The model}
We consider the following $d=(4+n)$ dimensional Einstein-Yang-Mills (EYM) model with action
\begin{equation}
 S=\int d^{4+n} x \sqrt{-g_{(4+n)}} \left(\frac{R}{16\pi G_{(4+n)}} - \frac{1}{2} {\rm tr} \left(F_{MN} F^{MN}\right) \right)   \ \ , \ \ M,N=0,...,3+n
\end{equation}
with the field strength tensor 
$F_{MN} = F_{MN}^{a} T_a=\partial_M A_N - \partial_N A_M + i [A_M,A_N]$ and the non-abelian
gauge potential $A_{M} = A_{M}^a T_a$ with the infinitesimal generators of the Lie group $T_a$.
Note that the gravitational coupling
$G_{(4+n)}=M_{planck}^{-(2+n)}$ has dimension of $[{\rm energy}]^{-(2+n)}$.
The coupled system of equations are the Einstein equations
$R_{MN} - \frac{1}{2} g_{MN} R = 8\pi G_{(4+n)} T_{MN}$, $M,N=0,..,3+n$
with the Yang-Mills (YM) energy-momentum tensor
\begin{equation}
 T_{MN}=2 \ {\rm tr}\left(g^{PQ}F_{MP}F_{NQ} - \frac{1}{4} g_{MN} F_{PQ} F^{PQ}\right) 
\end{equation}
and the Yang-Mills equations
$\nabla_M F^{MN} + i [A_M,F^{MN}]=0$.
In the following, we will abbreviate $8\pi G_{(4+n)}=\kappa$. 

\section{(Hyper)Spherically symmetric solutions}
\subsection{Equations of motion}
The Ansatz for the metric in Schwarzschild-like coordinates 
reads \cite{bct}:
\begin{equation}
 ds^2 = -\sigma^2(r) N(r)dt^2 + N^{-1}(r) dr^2 +r^2 d\Omega_{2+n} \ \ , \ \ N(r)=1-\frac{c_n \kappa m(r)}{r^{1+n}}
\end{equation}
where $d\Omega_{2+n}$ denotes the $(2+n)$ dimensional angular element and $m(r)$ the mass function such that $m(\infty)$ corresponds to the ADM mass of the solution.
$c_{n}$ is a constant that is given
by $c_n=2^{(2+n)/2}$ for $n$ even and $c_n=2^{(3+n)/2}$ for $n$ odd \cite{bct,bcht}.
 
The corresponding Ansatz for the $SO(\bar{d})$ gauge fields reads \cite{bct,bcht}:
\begin{eqnarray}
 A_0=0 \ \ , \ \ A_i=\frac{w(r)-r}{r} \Sigma_{ij}^{(\pm)} \hat{x}_j \ \ , \ \
\Sigma_{ij}^{(\pm)} = -\frac{1}{4} \left(\frac{1\pm \Gamma_{\bar{d}+1}}{2}\right) [\Gamma_i, \Gamma_j]  
\end{eqnarray}
with $\Gamma$ denoting the $\bar{d}$-dimensional gamma matrices and $\bar{d}=4+n$ for
$n$ even and $\bar{d}=3+n$ for $n$ odd.
The equations of motion then read \cite{bct,bcht}:
\begin{equation}
\left(r^{n} \sigma N w'\right)' = (1+n) r^{-2+n} \sigma (w^2-1)w 
\end{equation} 
\begin{equation}  
m' = r^{n} \left(Nw'^2   + \frac{(1+n)}{2} \frac{(w^2-1)^2}{r^2}\right) \ \ \ \ , \ \ \ 
\frac{\sigma'}{\sigma} = c_n \kappa\frac{w'^2}{r} 
\end{equation}
where the prime denotes the derivative with respect to $r$.

To ensure asymptotic flatness and finite energy solutions, we require:
\begin{equation}
\label{bc1}
 \sigma(r=\infty)=1 \ \ , \ \ w(r=\infty)=\pm 1
\end{equation}
while the conditions at the regular black hole horizon $r=r_h$ read 
\begin{equation}
\label{bc2}
 N(r=r_h)=0 \ \ , \ \  \left.
\left(N' w' - (1+n)\frac{(w^2-1)w}{r^2}\right)\right\vert_{r=r_h}=0
\end{equation}
and we assume $\sigma(r=r_h) > 0$ and finite.

\subsection{A no-go result for $n \ge 1$}
In \cite{bhr}, it was shown that {\bf no} hyperspherically symmetric Einstein-Yang-Mills
black holes exist in $n=1$. Here, we will extend this result to $n >1$ and
also point out the special role of $n=0$, i.e. the standard 4-dimensional case,
in which Einstein-Yang-Mills black holes are known to exist \cite{coloured}.
Following the discussion in \cite{volkov}, we find 
\begin{equation}
 (m\sigma)' = r^{-2+n}\left( w'^2 r^2 + \frac{1+n}{2} (w^2-1)^2 \right) \sigma
\end{equation}
Integrating this between the horizon $r=r_h$ and $r=\infty$ -- and using the boundary conditions (\ref{bc1}), (\ref{bc2}) -- we find
\begin{equation}
\label{meq}
 m(\infty) - \left.\left(\frac{r^{1+n}}{c_n \kappa} \sigma\right)\right\vert_{r=r_h} =
\int\limits_{r_h}^{\infty} dr \ r^{-2+n} \left(w'^2r^2 + \frac{(1+n)}{2} (w^2-1)^2\right)\sigma  \end{equation}
If we now study the behaviour of this latter expression under  a rescaling of the radial
coordinate $r\rightarrow \lambda r$, we note that the integral term scales like
$\lambda^{1-n}$, while the boundary term at $r_h$ scales like $\lambda^{-(1+n)}$.
Since the equation for $\sigma$ can be integrated to give $\sigma(r)=\exp(-c_n \kappa \int\limits_r^{\infty} \frac{w'^2}{r} dr)$, $\sigma$ scales with 
$\kappa \rightarrow \lambda^2 \kappa$. The mass $m(\infty)$ should be stationary under such a rescaling, i.e.
should fulfill $\frac{d}{d\lambda} m(\infty)\vert_{\lambda=1}=0$. Denoting the rhs
of (\ref{meq}) by $I(r_h)$, we find
\begin{eqnarray}
\label{scaling}
 \frac{d}{d\lambda} m(\infty)\vert_{\lambda=1} = (1-n) I(r_h) & - & 2 c_n \kappa I(r_h) \int\limits_r^{\infty} \frac{w'^2}{r} dr  -  (1+n) \left.\left(\frac{r^{1+n}}{c_n \kappa} \sigma\right)\right\vert_{r=r_h} \nonumber \\
&-&2 \left.\left(r^{1+n}\sigma \int\limits_r^{\infty} \frac{w'^2}{r} dr\right)\right\vert_{r=r_h}
\end{eqnarray}
This expression is strictly negative for $n \ge  1$. Thus {\it no static, finite mass hyperspherically symmetric non-abelian black holes exist in space-times with one or more extra dimensions.} It is also apparent that $n=0$, i.e. the standard 4-dimensional
case has a special role since for this, the first term in (\ref{scaling}) is positive
and can cancel the negative terms. Thus, finite mass black holes are possible. These solutions
have been extensively discussed in the literature \cite{coloured}.

\subsection{Finite mass black holes in generalized EYM models}
The idea here is to add additional terms from the gravitational or Yang-Mills hierarchy in order to circumvent the argument given in the previous section. The terms from the Yang-Mills
hierarchy are $F^2(2p)$, $p=1,2,3...$, where $F(2p)$ is the totally antisymmetrized
product of the Yang-Mills field strength tensor and reads \cite{bct,bcht}:
\begin{equation}
 F(2p)=F_{[\mu_1\mu_2} F_{\mu_3\mu_4}....F_{\mu_{2p-1}\mu_{2p}]}
\end{equation}
Under $r\rightarrow \lambda r$, these terms scale like $\lambda^{4p-n-3}$. As long as
$4p > n+3$ positive terms would be added to the rhs of (\ref{scaling}) and static, finite
mass, hyperspherically symmetric non-abelian black holes are possible. These solutions
have been constructed numerically in \cite{bcht} for $d=5$ and an additional $F(4)$ term. 
Let us summarize their main properties here: they have finite mass, but are not uniquely characterized by their mass
since for a specific range of parameters more than one solution with the same mass
can exist. Hence, in analogy to the 4-dimensional case existing
uniqueness theorems for static, hyperspherically symmetric, asymptotically flat
black holes do {\it not} generalize to theories containing Yang-Mills fields.
For a fixed value of the gravitational coupling, black hole solutions can only exist if their horizon radii are not ``too large'', i.e.
a maximal value of the horizon radius parameter $r_h$, $r_h^{(max)}$ exists. It turns
out that in the interval $r_h \in [r_h^{(cr,1)}:r_h^{(max)}]$ two black hole solutions exist.
In contrast to the 4-dimensional case, no solutions with more than
one node of the gauge field function were found.

\section{Non-abelian black strings}
We focus on the case $n=1$ here since most studies have been done for this case.
Thus, we want to study black objects with horizon topology $S^2\times S^1$, where we assume
the solutions to be translationally invariant in the extra coordinate direction $x^4$, i.e. $\frac{\partial}{\partial x^4}$ to be Killing. The Ansatz for the metric with $\mu,\nu,\rho=0,1,2,3$ then reads
\begin{displaymath}
 ds^2 = e^{-2/\sqrt{3}\psi(x^{\rho})} g^{(4)}_{\mu\nu}(x^{\rho}) dx^{\mu} dx^{\nu} + e^{4\sqrt{3}\psi(x^{\rho})}\left( dx^4 + 2{\cal W}_{\mu}(x^{\rho})dx^{\mu}\right)^2
\end{displaymath}
where $g^{(4)}_{\mu\nu}$ describes the 4-dimensional metric. The Ansatz for the gauge fields is $A_M dx^M= A_{\mu}(x^{\rho}) dx^{\mu} + A_4(x^{\rho})dx^4$ 
with $\mu,\nu,\rho=0,1,2,3$, where in the following we choose $x^{\rho}\in \{r,\theta,\varphi\}$, where $r$, $\theta$, $\varphi$ are isotropic coordinates.
We discuss deformed and undeformed non-abelian black strings here.
Undeformed refers to the fact that the solutions depend only on the 4-dimensional
radial coordinate $r$, i.e. are spherically symmetric for constant $x_4$, while
in the deformed case, the solutions can have a dependence on the 4-dimensional polar angle $\theta$. The latter solutions are then axially symmetric for constant $x_4$. The Ansatz in isotropic coordinates reads:
\begin{displaymath}
 g^{(4)}_{\mu\nu}dx^{\mu} dx^{\nu}=-f(r,\theta) dt^2 + \frac{q(r,\theta)}{f(r,\theta)}\left(dr^2 + r^2 d\theta^2 \right)
+ \frac{l(r,\theta)}{f(r,\theta)}r^2\sin^2\theta d\varphi^2
\end{displaymath}
for the 4-dimensional metric 
and 
$\psi=\psi(r,\theta)$ and  ${\cal W}_{\mu}=J(r,\theta)\delta_{\mu}^{\varphi}$
for the remaining metric functions. The Ansatz for the gauge fields reads:
\begin{eqnarray*}
A_M dx^M = \frac{1}{2r}\left[\tau_{\varphi}^n\left(H_1 dr + (1-H_2) r d\theta\right)   
- n\left(\tau_r^n H_3 + \tau_{\theta}^n (1-H_4)\right) r\sin\theta d\varphi +
\left(H_5 \tau_r^n + H_6 \tau_{\theta}^n\right) r dx^4 \right]
\end{eqnarray*}
with $H_i=H_i(r,\theta)$, $i=1,...,6$ and $\tau_r^n$, $\tau_r^{\theta}$ and
$\tau_r^{\varphi}$ denoting the scalar product of the vector of Pauli matrices $(\tau_x,\tau_y,\tau_z)$ with the unit vectors $\vec{e}_r^{(n)}=(\sin\theta \cos n\varphi, \sin\theta \sin n\varphi, \cos\theta)$, $\vec{e}_{\theta}^{(n)}=(\cos\theta \cos n\varphi, \cos\theta \sin n\varphi, -\sin\theta)$, $\vec{e}_{\varphi}^{(n)}=(-\sin n\varphi, \cos n\varphi, 0)$, 
$n$ denotes the winding of the solution.
Undeformed black strings can be constructed for $n=1$ with
$H_1=H_3=H_6 = 0$, $H_2=H_4=K(r)$, $H_5=H_5(r)$ and  
$f = f(r)$, $\psi=\psi(r)$ and $q=l=l(r)$. These solutions can be interpreted as
non-abelian black hole inside the core of a {\it spherically symmetric} dilatonic monopole
extended trivially into the extra dimension $x^4$. Correspondingly, deformed black strings
consist of a non-abelian black hole inside the core of an {\it axially symmetric} dilatonic monopole extended trivially into the extra dimension.

As such, both undeformed and deformed non-abelian black strings exist only as long as the horizon radius is smaller than the
core size of the dilatonic monopole and as long as the gravitational coupling is not too large. If the gravitational coupling is too large, the monopoles that carry
the mini black holes in their core would cease to exist themselves.
In fact, fixing the gravitational coupling and varying the horizon radius, two branches
of solutions exist. The solutions on the first branch are smooth deformations
of the corresponding globally regular solution with vanishing horizon radius. The two branches
of solutions meet at a maximal value of the horizon radius.
For small gravitational coupling the second branch can be extended all the way back to vanishing horizon radius where it bifurcates with a globally regular solution \cite{bh1,bhr}.
For larger gravitational coupling, the second branch terminates into an embedded abelian
solution of Einstein-Maxwell dilaton type.

The stability of the studied solutions is an important point. We will not investigate this
here in detail, but just want to point out that the results in \cite{bhr} hint to the
fact that the solutions are {\it thermodynamically unstable}. In \cite{bhr} a plot
of the temperature $T$ and entropy $S$ of the black string solution as function of the
horizon radius is given. While $S$ is an increasing function of the horizon radius,
$T$ decreases with the horizon radius. Thus the specific heat $C=T\left(\frac{\partial S}{\partial T}\right) < 0$ and the non-abelian black strings are thermodynamically unstable. 

\section{Summary and Outlook}
In this paper, we have discussed properties of static non-abelian black holes and black strings in higher dimensions. An important result is the non-existence of static, hyperspherically symmetric  black hole solutions of the standard Einstein-Yang-Mills model in
one or more extra dimensions, which we demonstrated here for $n >1$.
A possibility is the addition of terms from the Yang-Mills and gravity hierarchy, which
lead to finite mass solutions \cite{bcht}, which however are not uniquely characterized
by their mass. 
If one considers solutions of the standard Einstein-Yang-Mills model in space-times with compact extra dimensions so-called non-abelian black strings are possible.
While in theories without non-abelian fields black strings correspond to lower-dimensional spherical black holes times a compact manifold, non-abelian black strings can
be deformed in the sense that the black strings correspond to lower-dimensional
{\it axially symmetric} black holes times a compact manifold. This is a new feature
that appears due to the presence of the non-abelian gauge fields. 

Reinterpretation of the numerical results in \cite{bhr} points to the fact that non-abelian black strings are thermodynamically unstable. Whether these solutions are also classically stable
is an open problem. It would also be interesting to study whether the Gubser-Mitra 
conjecture \cite{gum} that connects the thermodynamical stability with the classical stability
is valid in this case.



\begin{theacknowledgments}
BH would like to thank Y. Brihaye, A. Chakrabarti, R. Durrer, E. Radu and D.H. Tchrakian for
discussions. Special thanks to the organisers of the Spanish Relativity Meeting (ERE) 2008.
\end{theacknowledgments}



\bibliographystyle{aipproc}   

\begin{thebibliography}{9}
\bibitem{remark} Extremal solutions, i.e. solutions with
vanishing surface gravity, of Einstein-Maxwell theory 
don't necessarily belong to the Kerr-Newman family. An example are
the Majumdar-Papapetrou multi black hole solutions \cite{pm}.
\end{thebibliography}


\IfFileExists{\jobname.bbl}{}
 {\typeout{}
  \typeout{******************************************}
  \typeout{** Please run "bibtex \jobname" to optain}
  \typeout{** the bibliography and then re-run LaTeX}
  \typeout{** twice to fix the references!}
  \typeout{******************************************}
  \typeout{}
 }


\end{document}